# Impact of within-voxel heterogeneity in fibre geometry on spherical deconvolution

Ross Callaghan*, Daniel C. Alexander, Marco Palombo, Hui Zhang

Centre for Medical Image Computing and Department of Computer Science, University College London, London, UK

*Corresponding author: ross.callaghan.16@ucl.ac.uk; UCL Centre for Medical Image Computing, 90 High Holborn, London, WC1V 6LJ, UK

## Abstract

Axons in white matter have been shown to have varying geometries within a bundle using ex vivo imaging techniques, but what does this mean for diffusion MRI (dMRI) based spherical deconvolution (SD)? SD attempts to estimate the fibre orientation distribution function (fODF) by assuming a single dMRI fibre response function (FRF) for all white matter populations and deconvolving this FRF from the dMRI signal at each voxel to estimate the fODF. Variable fibre geometry within a bundle however suggests the FRF might not be constant even within a single voxel. We test what impact realistic fibre geometry has on SD by simulating the dMRI signal in a range of realistic white matter numerical phantoms, including synthetic phantoms and real axons segmented from electron microscopy. We demonstrate that variable fibre geometry leads to a variable FRF across axons and that estimation of the fibre orientation distribution function (fODF) can be affected by this variability. When a voxel contains variable fibre geometry, assuming a single FRF can lead to misestimation of the fODF, potentially resulting in further downstream errors in techniques such as tractography.

### Highlights

- Variable fibre geometry within a voxel leads to variable fibre response functions within the voxel

- More complex fibre geometry leads to a wider range of fibre responses within a voxel
- When a voxel contains variable fibre geometry, assuming a single response can lead to misestimation of the underlying fibre orientation distribution function

**Abbreviations**

CC: corpus callosum, ConFiG: contextual fibre growth, CSD: constrained spherical deconvolution, dMRI: diffusion MRI, EM: electron microscopy, fODF: fibre orientation distribution function, FRF: fibre response function, HCP: Human Connectome Project, NODDI: neurite orientation dispersion and density imaging, PDF: probability density function, SD: spherical deconvolution, SH: spherical harmonic, SNR: signal-to-noise ratio, WM: white matter

# 1 Introduction

Diffusion weighted magnetic resonance imaging (dMRI) has been widely used to probe the structure and organisation of brain tissue, with one particular area of focus being the estimation of the orientational distribution of neuronal fibres in a voxel. This fibre orientation distribution function (fODF) is particularly interesting since it is used in tractography techniques to probe the structural connectivity of the brain which is important in many clinical and basic neuroscience studies (Catani and Thiebaut de Schotten, 2013; Dell'Acqua and Tournier, 2019; Johansen-Berg and Behrens, 2006). Whilst tractography has found many uses, there remain a number of challenges to the technique, including typically generating a large number of false positive connections (Jbabdi and Johansen-Berg, 2011; Maier-Hein et al., 2017; Schilling et al., 2019a; Thomas et al., 2014). One potential source of these issues could be due to difficulties in reliably estimating the fODF, where minor differences in fODF can lead to large differences in the tractograms created (Schilling et al., 2019a, 2019b). Accurate and reliable estimation of the fODF is therefore important to improve the accuracy of tractography techniques.

Many techniques have been developed for estimating the fODF, of which perhaps the most prominent are based on spherical deconvolution (SD). While there are a variety of spherical deconvolution methods, the central principle is the same - the diffusion weighted signal as a function of the azimuthal ($\phi$) and elevation ($\theta$) angles is modelled as a spherical convolution of the fODF, $F(\theta, \phi)$, with a kernel (called the fibre response function (FRF)), $R(\theta)$, the typical diffusion weighted signal from a single fibre population estimated *a priori*. By estimating an FRF from voxels where the signals are deemed typical of a single coherent fibre population, the fODF is determined by deconvolving this FRF from the signal. Implicit in this formulation is an assumption that one common FRF is shared across all fibre populations in the white matter (WM). Recently, some works have challenged this assumption, for instance (Schilling et al., 2019b) use known fODFs from histology to estimate the FRF in different WM regions, showing that the FRF does indeed vary across the WM and that this variation does affect the estimated fODF and tractography results.

Indeed, while the FRF is typically taken to represent the typical response from a fibre population on a voxel level, the way that the convolution is defined mathematically requires that it be identical across individual fibre populations within a voxel. In fact, recent works using electron microscopy (EM) and high-resolution x-ray imaging to investigate WM axonal morphology show that axons within a voxel have different shapes (Abdollahzadeh et al., 2019; Andersson et al., 2020; Lee et al., 2019), with varying diameters along their length and non-straight trajectories. It is reasonable, therefore, to propose that this heterogeneity in fibre geometry could lead to different fibres having different responses. This may lead to misestimation of the fODF when assuming a single FRF for all fibres, which can have downstream consequences for techniques including tractography.

A related factor arising from the convolution is that an assumption is made that there is no exchange between fibres or, equivalently, diffusion in different directions using this representation. In essence, this means that the fibres are implicitly assumed to be perfectly straight and pointing a given direction since any deviation from straight (i.e. curved or undulating fibres) would introduce directions that are connected, violating the non-exchange assumption. Additionally, this means that it is hard to reconcile the effect of tissue compartments which contribute to multiple directions, such as isotropically restricted compartments or the extracellular space, with this formulation. Under some experimental conditions, such as those used in current clinical applications, these effects may not be negligible and may affect subsequent techniques such as tractography.

In this work we investigate what effect, if any, violation of these assumptions introduced by within-voxel heterogeneity in axonal morphologies has on SD techniques. We use Contextual Fibre Growth (ConFiG) (Callaghan et al., 2020), our recently developed white matter numerical phantom generator capable of generating realistic WM morphology to investigate this in controllable environments, as well as real digital tissues reconstructed from EM (Lee et al., 2019) to test a limited sample of real tissue. Firstly, we investigate how microscopic variations in fibre geometry affect the diffusion within each fibre and whether the dMRI signal from each fibre is the same. We further evaluate what effect this has on fODF estimates by calculating them using FRFs representing the variable responses present in a voxel.

The rest of this paper is organised as follows: Section 2 describes the experiments performed to probe the assumptions outlined above, Section 3 presents the results and Sections 4&5 summarise the contributions and discuss future work.

# 2 Method

In order to test the impact of fibre geometry heterogeneity on the dMRI signal per-fibre and how any variability in response may affect fODF estimation. Experiments were performed with a range of numerical phantoms generated using ConFiG and reconstructed from EM (Lee et al., 2019), using an acquisition typical of SD applications.

Two primary experiments were conducted:
1. **Per-fibre response heterogeneity** - To investigate the impact of fibre geometry heterogeneity on the dMRI signal per-fibre
2. **Impact on fODF estimation** – To investigate what impact variation in the FRF can have on fODF estimation

In this section, we begin by describing how the phantoms were generated and gold-standard fODFs extracted and quantified, before describing the dMRI simulation experiments that were performed to investigate the impact of microscopic structural variability.

## 2.1 Phantom Generation

In order to test SD techniques in realistic geometries, a set of digital-tissue phantoms were generated to represent a range of WM tissue configurations:

- A single bundle of fibres generated by ConFiG with varying amounts of orientation dispersion:
    - Watson distributed (Mardia and Jupp, 2008), $\kappa = 2$ – highly dispersed
    - Watson distributed, $\kappa = 6$ – typical WM dispersion
    - Watson distributed, $\kappa = 100$ – highly coherent
- Crossing bundles of fibres generated by ConFiG
    - Two perpendicular bundles
    - Three perpendicular bundles
- Real fibres from mouse corpus callosum (CC) reconstructed from EM (Lee et al., 2019)

In the case of the single bundle phantoms, a low $\kappa$ means high orientation dispersion, so phantoms with a lower $\kappa$ were expected to have more complex morphology since higher OD means that they must grow around one another more to avoid intersections. A typical $\kappa$, estimated using NODDI (Zhang et al., 2012), for the corpus callosum of a healthy Human Connectome Project (HCP) (Stamatios N. Sotiropoulos et al., 2013; Van Essen et al., 2012) subject is $\kappa \sim 6$ (Callaghan et al., 2020) Since the CC is expected to contain some of the most coherent fibre bundles in the brain, $\kappa \sim 6$ will be towards the lower end of OD (higher end of $\kappa$) *in vivo*. Crossing bundle phantoms were generated by using starting and target points arranged into two- or three-crossing bundles and grown using ConFiG to generate complex phantoms with interleaved fibres.

### 2.1.1 Real WM fibres from EM

To simulate diffusion in real axons, 3D meshes were generated from WM axon segmentations from EM of mouse corpus callosum presented by (Lee et al., 2019). The axonal segmentations are provided in the NIfTI format, a volumetric format. In order to convert these into surface meshes for dMRI simulation, the `isosurface` function in MATLAB was used, however this produces meshes with some artifacts such as loose surfaces inside the fibres. In order to account for this, a further mesh refinement procedure was developed using the shrinkwrap feature in Blender to create a smooth, closed surface mesh around each fibre.

### 2.1.2 Gold standard fODF extraction from microstructure

In order to generate a ground truth to evaluate fODFs estimated from the simulated dMRI signals, a gold standard fODF was estimated directly from the WM numerical phantom meshes. As an attempt to generate a microstructural fODF comparable to that estimated from the simulated dMRI based fODFs, the microstructural fODF was calculated using the assumption of one direction per fibre, namely the direction connecting the endpoints of the fibre that would subsequently be used to align each fibre to the z-axis (see Figure 2).

A triangulated unit sphere was used to store this fODF, with each triangle in the sphere storing the number of fibres whose direction went through that triangle scaled by the volume of each fibre, as illustrated in Figure 1. In order to compare this fODF to those calculated using SD from dMRI, the microstructural fODF was expanded in spherical harmonics (SHs). A spherical function $f(\theta, \phi)$, can be expressed in terms of spherical harmonics as:

$$f(\theta, \phi) = \sum_{l=0}^{l_{max}} \sum_{m=-l}^{l} c_l^m \, Y_l^m(\theta, \phi) \, , \qquad (1)$$

where

$$c_l^m = \int_0^{2\pi} \int_0^{\pi} f\,(\theta,\phi) Y_l^{m*}(\theta,\phi) \sin(\theta) d\theta d\phi\,. \tag{2}$$

$Y_l^m$ are the so-called spherical harmonics of degree $l$ and order $m$ up to a maximum degree $l_{max}$ and $*$ denotes complex conjugation. In our case, $\theta$ and $\phi$ are discrete samples in the centre of each triangle in our unit sphere meaning one approach to finding $c_l^m$ is to turn the integral in into a summation. We adopt a more robust approach based on least squares (Alexander et al., 2002; Brechbühler et al., 1995) in which the spherical harmonics are re-indexed to have single index $j(l,m) = l^2 + l + m$. The discrete fODF values stored in each triangle are turned into a vector of length $n_{\mathrm{tri}}$, $[f] = \{f(\theta_i, \phi_i), i = 1, \dots, n_{tri}\}$ and an $n_{\mathrm{tri}} \times j(l_{\mathrm{max}}, l_{\mathrm{max}})$ matrix, $X$, constructed with elements $X_{i,j(l,m)} = Y_l^m(\theta_i, \phi_i)$. Essentially, $X$ maps the SH coefficients for each $l, m$ into amplitudes along each $\theta_i, \phi_i$. The $j(l_{\mathrm{max}}, l_{\mathrm{max}})$ vector of SH coefficients, [c] can then be found as

$$[\mathrm{c}] = (X^{*T} X)^{-1} X^{*T} [f]\,. \tag{3}$$

The number of coefficients in $[c]$ can be reduced since the fODF is real-valued and antipodally symmetric. Being real valued means that the SH coefficients exhibit conjugate symmetry (that is, $c_l^m = (-1)^m c_l^{-m*}$) and the antipodal symmetry means that all odd $m$ terms are 0 (Alexander et al., 2002; Tournier et al., 2004). In the end, this means that $[c]$ has $(l_{max} + 1)(l_{max} + 2)/2$ elements.

Each fODF was normalised such that

$$\int_0^{2\pi} \int_0^{\pi} f\,(\theta,\phi) \sin(\theta) d\theta d\phi = 1\,, \tag{4}$$

to ensure that the fODF is a probability density function (PDF), describing the probability of fibre pointing in a given unit of solid angle.

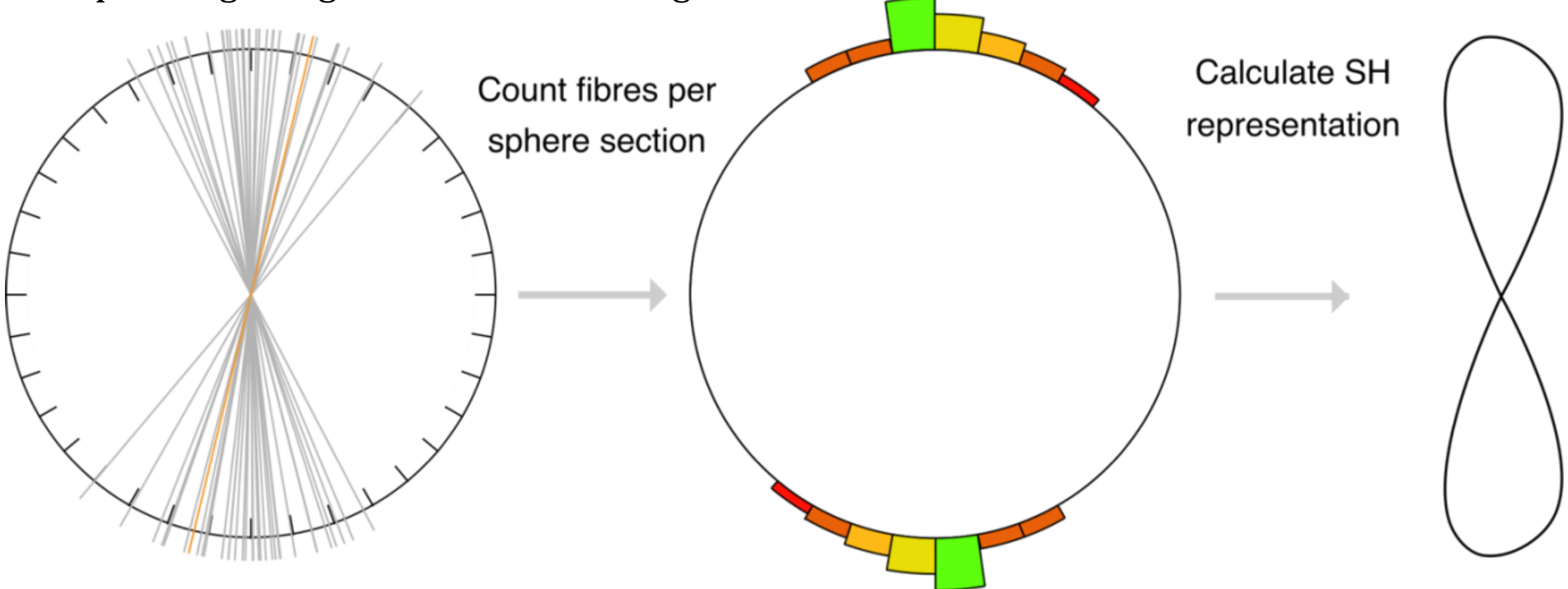

*Figure 1 Gold standard fODF estimation from microstructure of WM numerical phantoms. Each fibre's main direction (line in left image, extracted as shown in Figure 2) is projected onto a sphere and a spherical harmonic representation calculated*

#### 2.1.3 fODF metrics

In order to quantify differences in fODFs, a series of metrics were extracted by fitting a Watson distribution to each single bundle fODF. The Watson distribution probability density function is defined as a function of a unit vector $\vec{x}$, parameterised by $\vec{\mu}$ and $\kappa$, the mean direction and concentration parameter respectively.

$$f(\vec{x}|\vec{\mu},\kappa) = C(\kappa)\exp(\kappa(\vec{\mu}\cdot\vec{x})^2)\,, \qquad (1)$$

where

$$C(\kappa) = M\left(\frac{1}{2},\frac{3}{2},\kappa\right)^{-1} \qquad (2)$$

is Kummer's confluent hypergeometric function which is a constant for any given $\kappa$. If we have a series of measured directions $\{x\} = \{\vec{x}_1, \vec{x}_2, \dots, \vec{x}_n\}$, we can determine the Watson best-fit parameters with maximum-likelihood estimation. The likelihood function can be computed as

$$L(\vec{\mu},\kappa|\{x\}) = \prod_{i=1}^{n} f\,(\vec{x}_i|\vec{\mu},\kappa)\,,$$
$$= \prod_{i=1}^{n} C\,(\kappa)\exp(\kappa(\vec{\mu}\cdot\vec{x}_i)^2)\,. \qquad (3)$$

In this experiment, $\{x\}$ is our set of measured fibre directions from the phantom, however since each fibre has a different volume, we need to create a weighted likelihood which weights larger fibres more heavily than small fibres. Each fibre should contribute to the overall distribution proportionally to its volume, so we assign each fibre a weight which is its volume, $v_i$, normalised by the overall fibre volume $V = \sum_i v_i$ , i.e. the weight for each fibre is $w_i = {}^{v_i}/_{V}$.

In order to incorporate this into the likelihood, we treat each fibre contributes as if it contributes $w_i$ times so the overall likelihood, so Eq. 3 becomes

$$L(\vec{\mu},\kappa|\{x\}) = \prod_{i=0}^{n} [C(\kappa)\exp(\kappa(\vec{\mu}\cdot\vec{x}_i)^2)]^{w_i}. \qquad (4)$$

Taking the logarithm and exploiting logarithmic properties, this can be simplified to give the log-likelihood

$$l(\vec{\mu},\kappa|\{x\}) = \sum_{i=0}^{n} w_i\,[\log(C(\kappa)\,\exp(\kappa(\vec{\mu}\cdot\vec{x_i})^2))]\,,$$

$$= \ logC(\kappa) + \kappa\sum_{i=0}^{n} w_i(\vec{\mu}\cdot\vec{x_i})^2\,. \qquad (5)$$

This can be further simplified using the weighted scatter matrix, $\mathbf{T}(\vec{w},\mathbf{x}) = \sum_{i=0}^{n} w_i\,\vec{x_i}\otimes\vec{x_i}$ where $\otimes$ denotes the outer product, to give

In order maximise the log-likelihood, we follow the approach in (Mardia and Jupp, 2008), to directly estimate $\vec{\mu}$ using the scatter matrix, $\mathbf{T}(\vec{w},\mathbf{x}) = \sum_{i=0}^{n} w_i\,\vec{x_i}\otimes\vec{x_i}$ , where $\otimes$ denotes the outer product. The maximum likelihood estimate of $\vec{\mu}$ is then found by taking the eigenvector of $\mathbf{T}$ with the largest eigenvalue (Mardia and Jupp, 2008). $\kappa$ is then estimated with an iterative grid search evaluating Equation 5 for a range of possible $\kappa \in [0,100]$.

Additionally, this approach is used to quantify dMRI derived fODFs by sampling uniform directions on the sphere and taking the weight, $w_i$, for each direction to be the normalised fODF amplitude in that direction.

## 2.2 Experiments

### 2.2.1 Per-fibre response heterogeneity (Experiment 1)

We test to what extent variable fibre geometry results in variable fibre responses by simulating the intracellular dMRI signal from the phantoms described in Section 2.1 following the procedure outlined in Figure 2.

As mentioned in the Introduction, SD assumes that the overall voxel signal is the sum of signals from the constituent fibre directions. This requires no exchange between different fibre directions which makes the extracellular space difficult to account for. Indeed, in practice the signal is often assumed to come from the intra-axonal space alone to aid in interpretation of the resulting fODF (Raffelt et al., 2012, 2017). Therefore, to test SD under the assumptions inherent in the model, only the intra-axonal signal was simulated.

To create a periodic intra-axonal space for simulation, each fibre was rotated to be aligned with the $z$-axis and then extended with a reflected copy as in (H. Lee et al., 2020). The rotation matrix used to align the fibre with $z$ was stored so that signals could be rotated back into the dispersed directions to generate an overall voxel signal. In the case of EM fibres, only axons that were longer than 18 μm(chosen to be slightly shorter than the height of the EM volume in the principle fibre direction) before extension were used in simulation to remove very short fibres that had been segmented at the edge of the volume.

Diffusion MRI signals were simulated from ConFiG phantoms using the Camino dMRI simulator (Cook et al., 2006; Hall and Alexander, 2009) to perform the experiments described below. For all experiments a bulk diffusivity D = 2.0μm$^2$/ms was used in accordance with similar Monte Carlo experiments (Dhital et al., 2019; Hall and Alexander, 2009; H.-H. Lee et al., 2020; Palombo et al., 2020). Standard Camino periodic boundaries were used in which copies of the phantom are tiled in the x-, y- and z-directions to create an effectively infinite, but periodic, simulation domain.  (Panagiotaki et al., 2010).

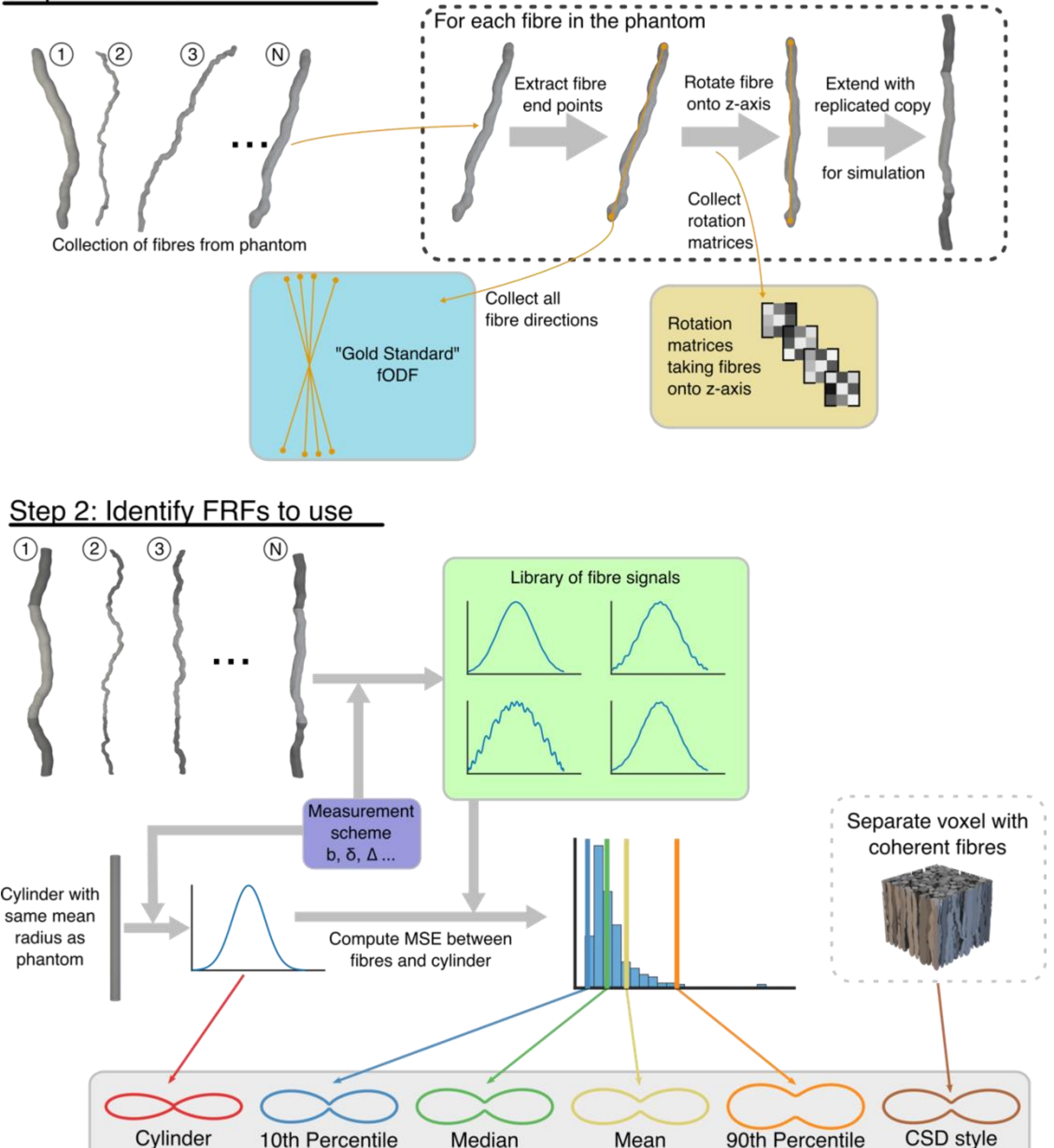

*Figure 2 Steps 1 and 2 in the experiment pipeline. Step 1 processes fibre meshes to prepare them for simulation as well as to extract gold standard fODF metrics. Each fibre is rotated to align with the z-axis and extended with a reflected copy for simulation as in (Callaghan et al., 2020; H. Lee et al., 2020). The main direction of each fibre as well as the rotation matrix used to align it to z are stored for later use. Step 2 simulates the dMRI signal in each fibre, taking the signal only from the original (not replicated) copy of the fibre and computes the distance*

*between each fibre's signal and a cylinder signal to extract a set of FRFs. This gives a measure of variability in fibre responses in a voxel.*

For each fibre 10,000 spins were initialised uniformly within the intra-axonal space and the simulations were performed using 5000 timesteps. Each phantom had $\sim 300$ fibres giving $\sim 3 \times 10^6$ spins in total per phantom. These settings were confirmed to be adequate by comparing to a set of test simulations performed using $10^5$ spins per fibre and $10^4$ timesteps. The measurement parameters were $\Delta = 28$ms, $\delta = 24$ms, $b = 1000, 2000, 3000$ s/mm$^2$ and 256 equidistributed gradient directions (Saff and Kuijlaars, 1997) at each shell. This gives a diffusion time $d_t$=20 ms, chosen so that the diffusion length scale ($\sqrt{2Dd_t} \approx 8$µm at $D = 2.0$ µm$^2$/ms) is small relative to the axon length ($\geq$ 18µm). Additionally, these settings give $\mathrm{G} = 60$ mT/m at $\mathrm{b} = 3000$ s/mm$^2$, a feasible gradient strength on a high-end clinical system.

To compare to the collection-of-straight-fibres assumption implicit in SD techniques, an infinite cylinder representing each fibre was generated using the endpoints of each fibre to give the direction and the mean radius of the fibre as the cylinder radius. For simulation, each cylinder was aligned with the $z$-axis similarly to the ConFiG fibres so that everything was in the same space to compare the signals. The same measurement scheme was simulated in each cylinder in order to compare to the ConFiG fibres.

Since the individual axons have been aligned with the $z$-axis, the signals from each fibre can be directly compared with one another as the gradient directions are aligned with respect to each fibre. To demonstrate the variability in dMRI response, the median, 10th and 90th percentile responses were found (in terms of mean squared difference between fibre and cylinder responses). Additionally, to relate the type and size of morphological variation to the signal changes, the microscopic orientation dispersion (Brabec et al., 2019) ($\mu OD$), a measure of undulation, and the coefficient of variation of diameter ($CV$), a measure of beading (H. Lee et al., 2020), was calculated for each fibre.

Throughout this work, SH representations of signals are used. The MATLAB implementation of constrained spherical deconvolution (CSD) (Tournier et al., 2007) available from (https://github.com/jdtournier/csd) is used to calculate SH decompositions of signals. This is the same technique as used in popular dMRI tractography tool MRtrix3 (Tournier et al., 2019) and follows the procedure outlined in Section 2.1.2 for SH decomposition of the dMRI signal.

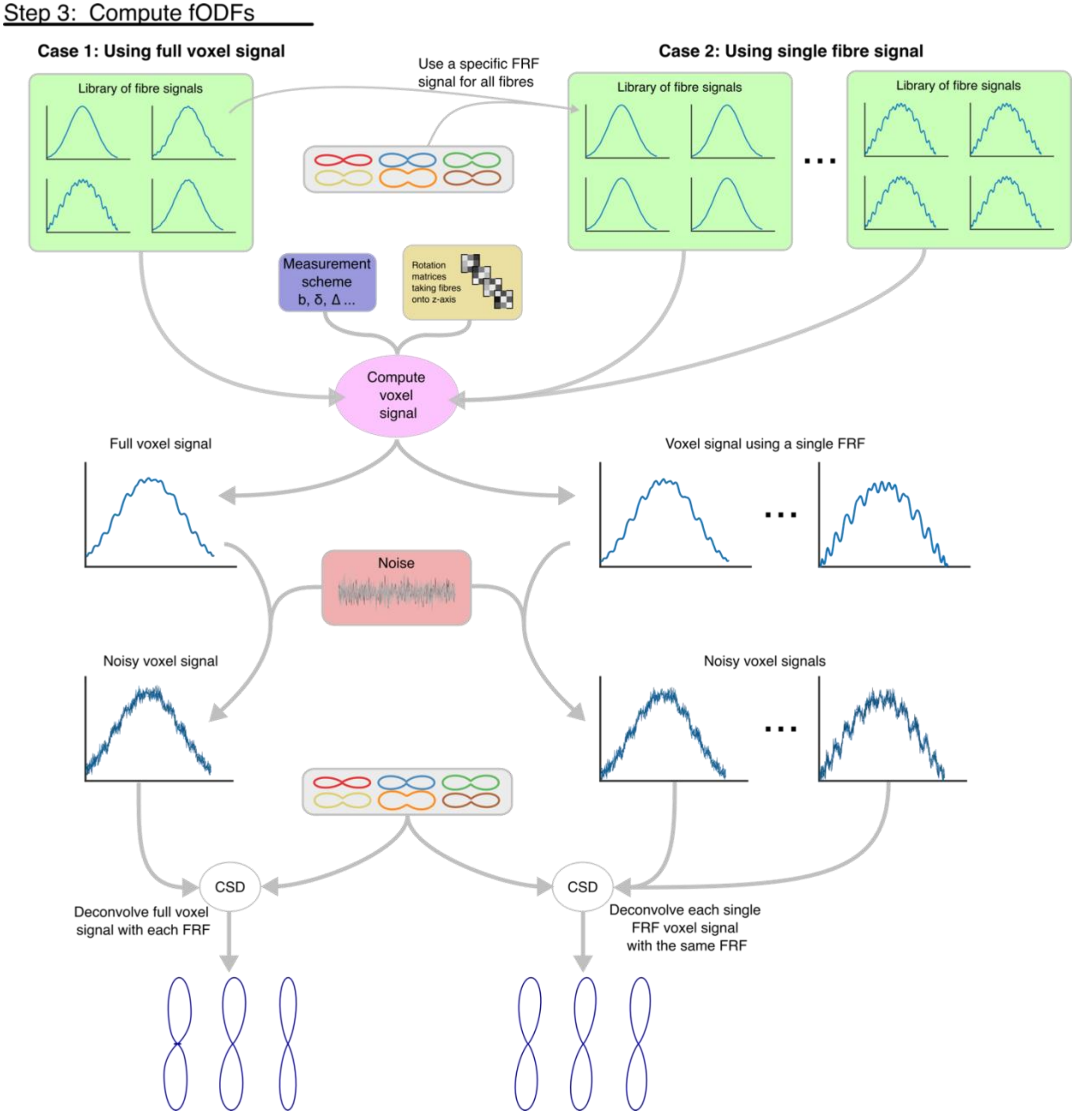

*Figure 3 The third step in the experiment is to compute voxel signals and perform CSD. There are two strands to the experiment, one using the full voxel signal from each fibre's individual signal and one using a single signal for all fibres, representing the ideal case in which there is a single true FRF. In each case the voxel signal is computed by combining all fibre signals*

*rotated into their original directions and then noise is added. The various FRFs used in the experiments are then deconvolved with the signal using CSD to estimate the fODF.*

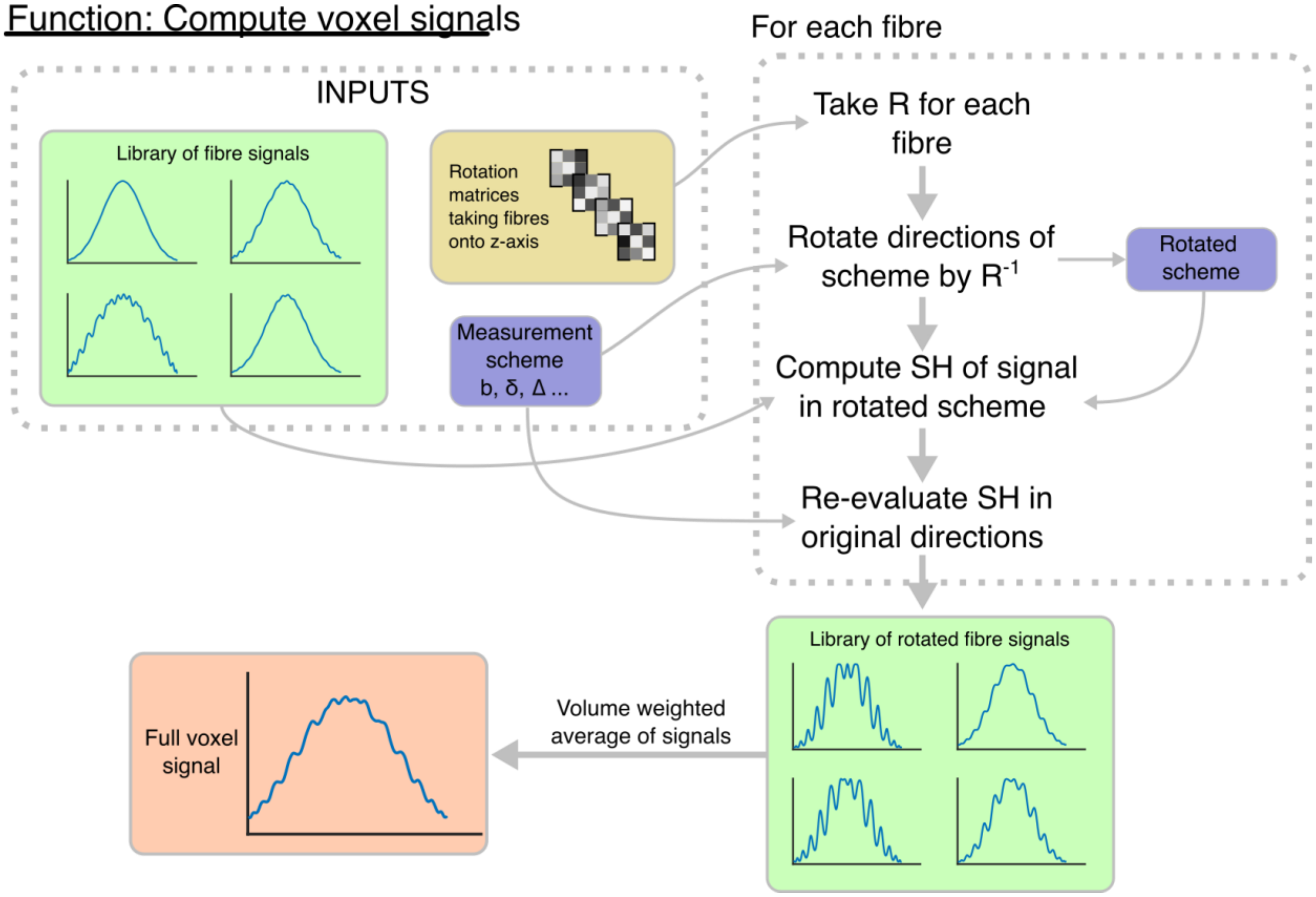

*Figure 4 Illustration of the method to produce voxel signals. Given a library of fibre signals aligned with z, the rotation matrices taking each fibre onto z and the measurement scheme, each fibre's signal is evaluated in SH in a rotated scheme and then these SH coefficients are re-evaluated in the original directions. Rotated signals are then combined, weighted by the volume of each fibre, to give the total voxel signal.*

### 2.2.2 Impact on fODF estimation (Experiment 2)

To investigate the impact on the fODF of assuming a single fibre response per voxel, we compared fODF estimates from CSD using six different FRFs per phantom derived from Experiment 1: the 10th, 50th and 90th percentile FRF (representing the spread of responses), the volume-weighted mean signal (representing a best case CSD scenario), a cylinder (representing an ideal stick-like response) and the $\kappa = 100$ voxel signal (representing a typical CSD-like approach of estimating the FRF in a region of coherent fibres). These FRFs are illustrated in Figure 2.

For each FRF, the fODF was estimated in two scenarios per phantom, illustrated in Figure 3:

1. Full voxel signal: The voxel signal contains each fibre's signal as simulated. This represents the realistic scenario where each fibre may have a different signal.
2. Single FRF signal: The voxel signal comes from using the same FRF signal per fibre. This represents the ideal CSD scenario in which the signal truly comes from a single FRF.

The approach to generating the voxel signals is illustrated in Figure 4, achieved by rotating each fibre's signal onto the original fibre direction and weighting by fibre volume. To investigate the impact of FRF variation relative to noise, the experiment was repeated for 500 Rician noise instances at 30 SNR. The fODF estimated using CSD is not a true PDF as it does not integrate to one, so throughout this work the fODF from CSD is normalised as outlined in Section 2.1.2.

# 3 Results

## 3.1 Per-fibre response heterogeneity

Variations in intra-voxel fibre geometries are present in real fibres and ConFiG phantoms as demonstrated in Figure 5 which shows each digital phantom alongside the fibres which give the median, 10th and 90th percentile response.

Under the experimental conditions investigated, this morphological variation in the fibres causes the dMRI signal response per-fibre to vary as can be seen in Figure 5, which shows the mean, median, 10th and 90th percentile signals across all fibres in each phantom at b=3000s/mm$^2$ and 30 SNR. The variation in the response function depends on the complexity of the fibre arrangement, with the most complex three-crossing bundle arrangement leading to the largest variation in response functions.

This variation in the FRF is seen across each $b$-value from 1000 to 3000s/mm$^2$ as demonstrated in Figure 6 for the $\kappa = 2$, three-crossing and EM fibre phantoms, chosen since these display the most variation for each phantom category. Here $SNR = \infty$ to isolate the effect of $b$ from noise.

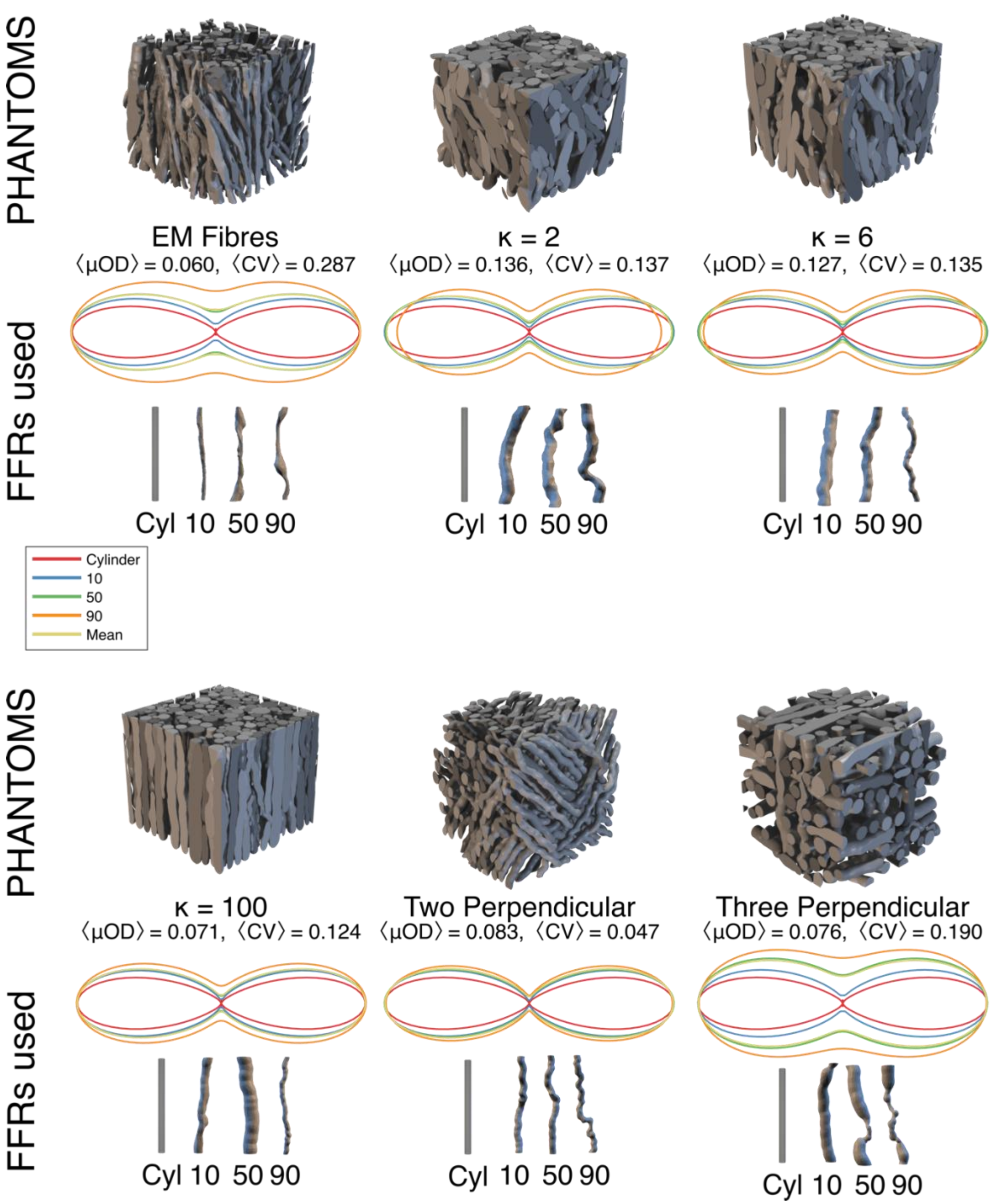

*Figure 5 Variability in fibre responses within a voxel at b=3000 s/mm$^2$ at SNR=30 along with geometrical variation in fibres responsible for median, 10$^{th}$ and 90$^{th}$ percentile response.*

## 3.2 Impact on fODF estimation

The variation in the FRF for each fibre leads to a variation in the estimated fODF as seen in Figure 7-Figure 9. Again, the magnitude of differences in fODF tends to depend on the complexity of the fibre arrangements since the more complex arrangements have more variation in the FRF. Generally, the fODF calculated from SD picks out the correct main peak direction that is seen in the gold standard fODF from the microstructure, with differences in the overall peak amplitude and shape. Notable exceptions to this occur for some phantoms when using the cylinder FRF, such as EM fibres and three-crossing bundles in which fODF peaks are estimated which are not present in the gold standard fODFs.

These qualitative differences in fODF produced with differing FRFs lead to quantitative differences as seen in Figure 9 for the single bundle phantoms. The more stick-like FRFs lead to an underestimation of $\kappa$ (too much orientation dispersion in the fODF) while FRFs which are more representative of the overall fibre population (mean and median) perform better. In the ideal single FRF voxel signal case, in which the signal truly comes from a single FRF, the fODF is generally estimated well with the main exception being a misestimation of peak direction for $\kappa$=2 and 6 with the 90th percentile FRF.

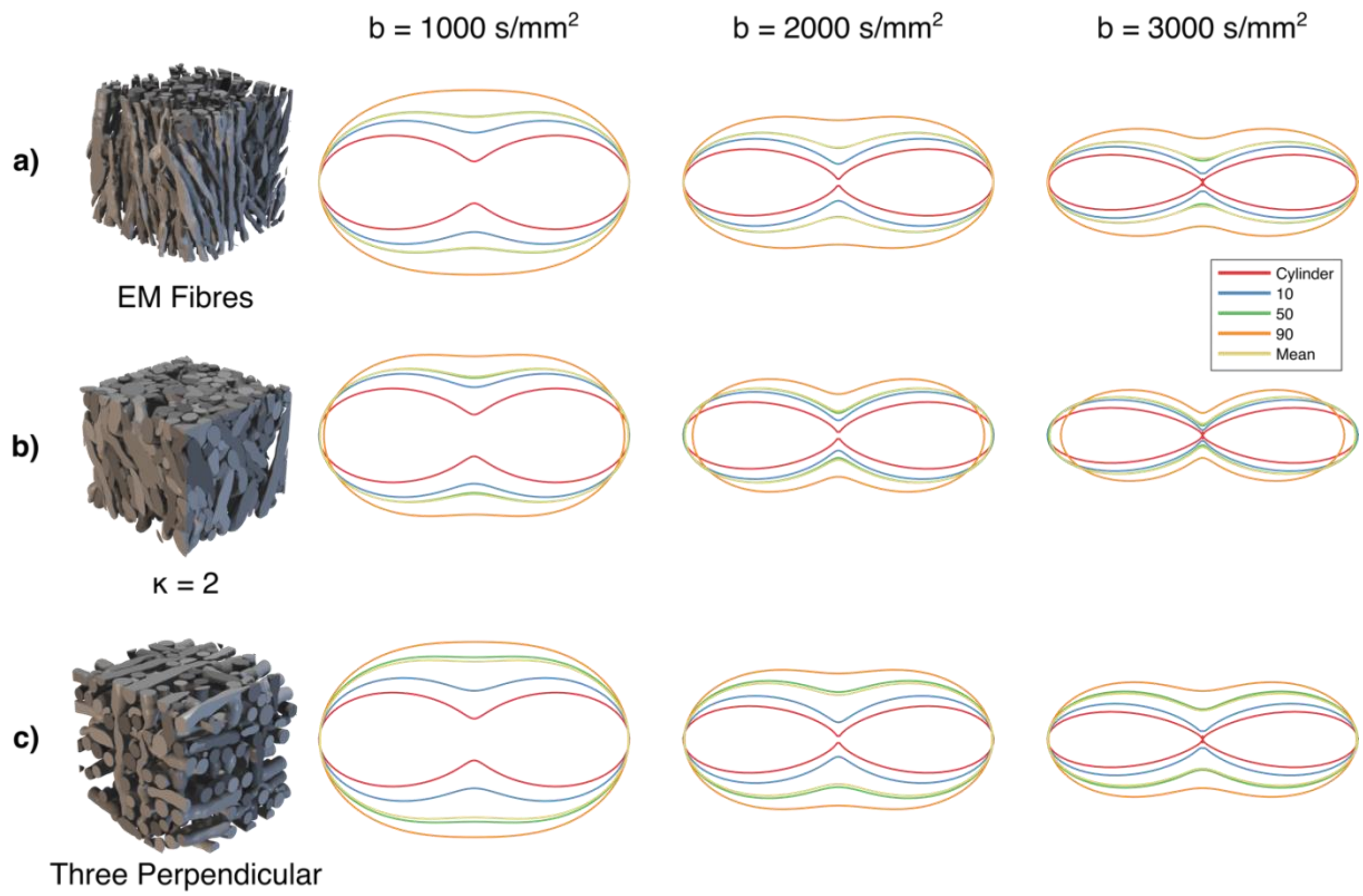

*Figure 6 Per-fibre response function at* $b = 1000, 2000, 3000\ s/mm^2$ *(left-to-right) for (a) the EM fibres, (b) the κ = 2 phantom and (c) the three-crossing phantom.*

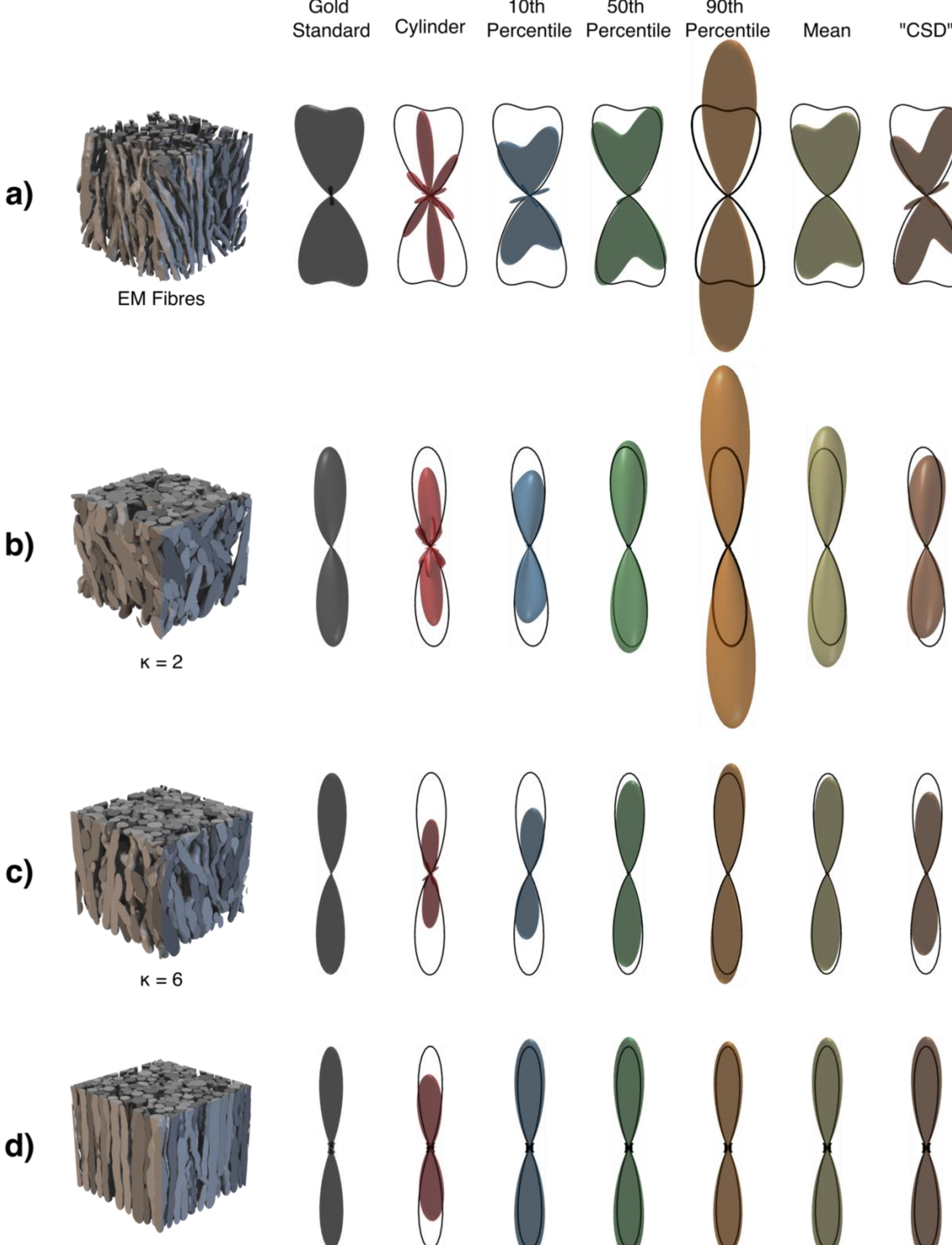

*Figure 7 Variation in fODF estimated using the range of FRFs in Experiment 2 at b = 3000 s/mm2 for the single fibre bundle voxels.*

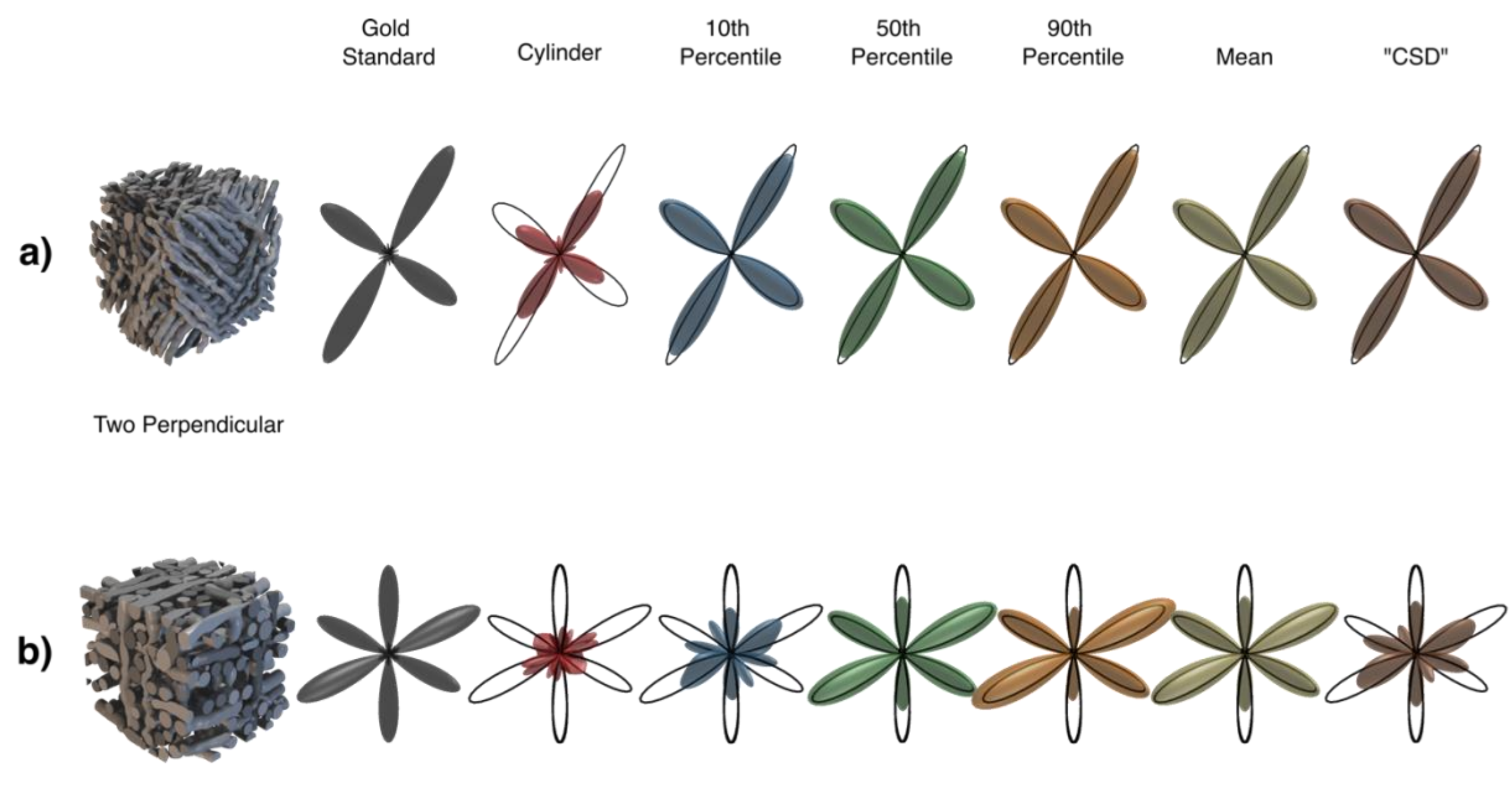

*Figure 8 Variation in fODF estimated using the range of FRFs in Experiment 2 at b = 3000 s/mm2 for the crossing fibre bundle voxels.*

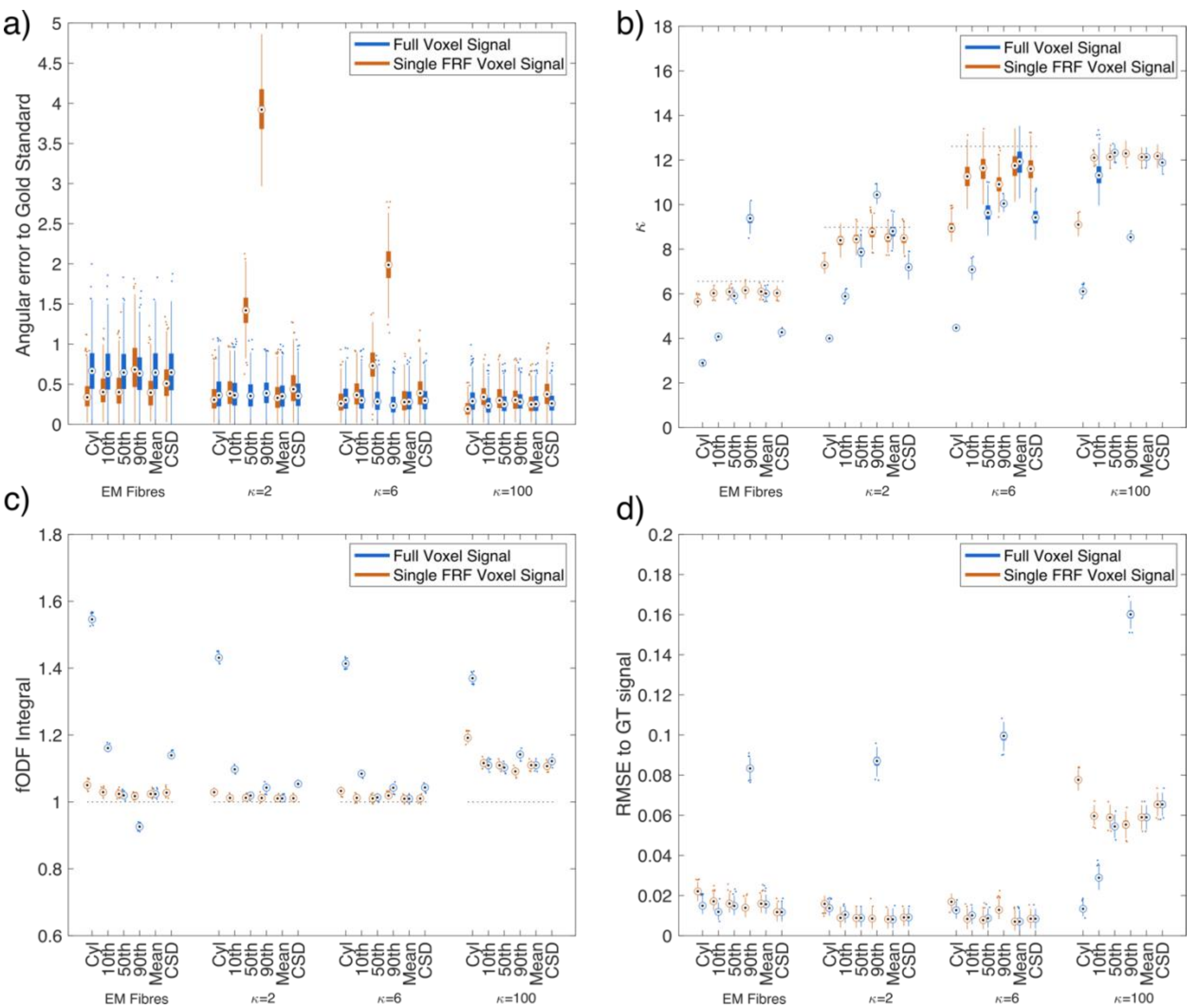

*Figure 9 Variation in fODF properties when estimated using different FRFs. a) the angular error between the main peak in gold standard fODF and that of the fODFs estimated using CSD, b) κ estimated from the fODF (gold standard value is dashed line for each fODF), c) fODF integral for each fODF estimated and d) the RMSE to the ground truth signal.*

# 4 Discussion

The microscopic variations in fibre morphology challenge the assumption in SD techniques that there exists a unique and shared FRF even within a single voxel. Here we have used dMRI simulations to demonstrate that variations in individual axonal morphology do indeed lead to different response functions per-fibre which in turn can have a knock-on effect on SD techniques to estimate the fODF.

As demonstrated in Figure 5 and Figure 6, the response function can vary substantially across different fibres, particularly in complex fibre arrangements such as the three-crossing bundles scenario. Indeed, the variation in responses in fibres reconstructed from real EM images of WM is large, similar to that of the ConFiG three-crossing simulations, and generally larger than ConFiG phantoms containing a single fibre bundle. This suggests that there is in fact more variation in real axons than in the axons generated by ConFiG, even in simple arrangements of a single bundle of fibres. This is something that may be used to inform future versions of ConFiG to generate more realistic phantoms.

In the main, the largest variation in the per-fibre response, as well as the largest difference between cylinders and realistic axons happens in the axial direction which is to be expected. Even with diameter variations and undulation, the radial diffusion is still strongly restricted under the assumption of no axonal permeability, while recent studies have shown that real axonal morphology causes time-dependent deviations from Gaussian diffusivity along the axial direction (H. Lee et al., 2020). Phantoms which show large amounts of beading (high $CV$ in EM fibres and three-crossing bundles) show the largest variability in response, while $\mu OD$ affects the response less, suggesting that fibre beading drives fibre response variability more than undulation. One exception to this in the $\kappa = 2$ case in which the 90th percentile fibre contains a large amount of undulation, leading to a reduction in radial signal at higher b-values (Figure 6).

Variations in the FRF have an impact on the estimated fODF as demonstrated in Figure 7-9. In the simplest fibre arrangements with low dispersion ($\kappa = 6, 100$, Figure 7c&d) the FRF variation is relatively small and the resulting fODF variation is small. In phantoms with more variable FRFs (EM Fibres Figure 7a, three-crossing Figure 8b), the resulting fODFs show more variability, sometimes identifying fewer (e.g. EM fibres 90th percentile) or more peaks (e.g. EM fibres cylinder, three-crossing CSD) than are present in the gold standard fODF and often producing fODFs with differently shaped peaks. This is significant because as shown by (Schilling et al., 2019b), even changes in fODF peak amplitudes without peak direction changes can have an impact on tractography results.

When constructing a voxel signal that truly comes from a single FRF and using that FRF to deconvolve, the fODF is generally recovered well as demonstrated in Figure 9. The main exception is the peak direction for κ=2 and 6 for the 90th percentile FRF, though this may be caused by the FRF itself containing some orientation dispersion, which makes the voxel signal more difficult to deconvolve as the assumption that each fibre contributes to a single direction is not valid.

Conversely, when using the full voxel signal, the fODF reconstructed depends on the FRF used as shown in Figure 9. This demonstrates that when the voxel signal truly comes from fibres with a mix of different responses, assuming a single FRF can lead to misestimation of the fODF. This effect is particularly pronounced when using an FRF that does not represent the fibre population well (Cylinder, 10th percentile, 90th percentile) and would be expected in these cases.

When using an FRF that more broadly represents the fibre population (mean and median), the fODF is generally well estimated, however it is worth noting that the typical CSD approach of taking the signal from a voxel of coherent fibres does not always perform well, producing wrongly shaped peaks for the EM fibre and three-crossing phantoms.

The main takeaway from these investigations is that within-voxel heterogeneity in fibre geometry leads to heterogeneity in the per-fibre response to the extent that using a single FRF in CSD cannot always accurately recover the underlying fODF. Some techniques account for some FRF variation voxel-to-voxel (Christiaens et al., 2017; Kaden et al., 2016b, 2016a), however the investigations presented here suggest that the FRF may vary even within a voxel. In spite of this, when using a FRF that broadly represents the population, such as the median fibre response, CSD is able to recover the fODF well, however the fact remains that currently there is no way to determine this representative FRF which means that SD techniques may suffer from this effect when prescribing a single FRF.

Additionally, these simulations lend further support to challenge the assumption that FRF is constant across the brain as differences in the mean FRF per phantom demonstrate that the overall FRF from different fibre arrangements will be different as a result of the different axonal morphology in each environment. Further, as demonstrated by the CSD-style fODF experiments, using a single FRF across different voxels can lead to misestimation of the estimated fODF, meaning that using the wrong FRF in different brain regions could have large impacts on fODF-based techniques such as tractography. For instance, this could potentially explain fODFs with many peaks in the gyral blade such as those seen in Figure 10, which appear similar to those seen in this study in Figure 7a when using an incorrect FRF and can lead to spurious fibres crossing the gyral blade in tractography. In this region, it is reasonable to expect complex fibre geometry (and therefore variable FRF) as axons spread out to reach their grey matter termini, however other factors may also influence this such as partial volume. In the absence of *in vivo* ground truth we cannot say for certain the cause of these fODFs, however this study suggests axonal morphology variations could be a contributing factor.

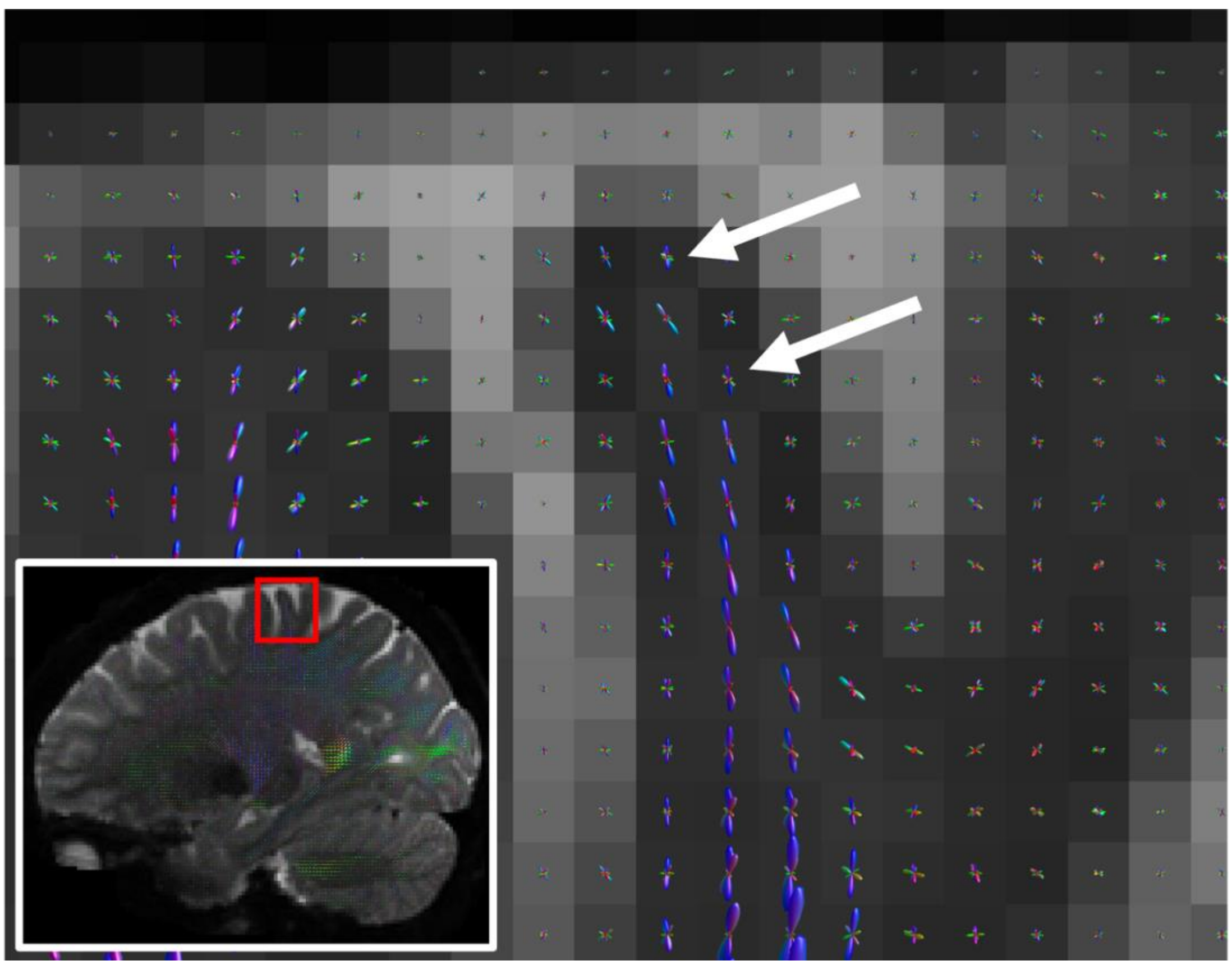

*Figure 10 An example of spurious peaks in the fODF estimated in the gyral blade of an HCP subject (S. N. Sotiropoulos et al., 2013; Stamatios N. Sotiropoulos et al., 2013)using MRTrix3 (Tournier et al., 2019). Fibres are expected to project towards the gyral crown but may end up with erroneous peaks in the fODF which can lead to errors in tractography (Cottaar et al., 2021; Van Essen et al., 2014).*

## 4.1 Limitations and future work

In this work we investigated how complex fibre morphology affects SD techniques using a dMRI scheme chosen to use clinically feasible gradient strength and duration, though it may be possible that other measurement schemes exhibit greater or lesser sensitivity to these effects. For instance (Yeh et al., 2010) have shown that the gradient pulse duration can impact fibre orientation estimation. Another factor to investigate is the impact of the diffusion time on the observed effects, since longer diffusion times can 'smooth out' these microscopic morphological variations as spins are able to diffuse further. At present, the diffusion time we can simulate is limited by the size of our phantoms, however work is ongoing to produce larger phantoms with ConFiG to allow us to study these effects at longer diffusion time.

Throughout this work we have referred to the fODF generated from the phantom microstructure as the 'gold standard' fODF in inverted commas. This is deliberate since it is not straightforward to define an fODF from microstructure that exactly corresponds to that from dMRI, in part due to the assumptions made in modelling the fODF from dMRI, which have been discussed here. Efforts have been made to make the two fODFs comparable in this work by defining the gold standard fODF using a single direction per fibre and normalising all fODFs to one.

The variability that is demonstrated in the per-fibre response in this experiment is suggested to arise due to the complexity of fibre morphology introduced due to complex fibre arrangements. This seems to be case, as shown in Figure 5, though the exact nature of the link is not known for certain since there are many sources of morphological complexity (undulation, beading, non-circular cross-sections, etc.) which could all contribute to these variations. Currently, ConFiG does not allow us to directly control such morphological features but instead the precise morphology generated is a product of the target fibre arrangement (orientation dispersion, density etc.) and the growth algorithm. Future work will aim to incorporate control over microscopic morphological features into ConFiG in order to isolate each of these effects to probe which morphological features have the largest impact on the FRF/fODF. Should these effects be understood, it may be possible to estimate them from the dMRI signal to improve the accuracy of SD techniques.

Another important consideration is that in this work we use CSD as in MRtrix3, however there are wide range of SD techniques for fODF estimation, each with slightly different derivations and assumptions. While this will affect the fODFs presented in this work, the per-fibre signals and compartmental signals presented do not rely on any SD model, so the FRF variations will impact any models which use the FRF.

It is worth pointing out that here we merely demonstrate that the response varies on a per-fibre basis, meaning that the concept of a fibre response function needs to be treated carefully. Most SD techniques estimate their FRF from averaging across a number of voxels, meaning that the FRF is an average single bundle response (including many fibres, extracellular space, other cells etc.) rather than purely a *fibre* response function. The variability in the per-fibre response may contribute less to the variable overall FRF seen across the brain in previous studies (Christiaens et al., 2020; Schilling et al., 2019b) than other factors such as the extracellular space but it should be a consideration.

It is also worth noting that this effect will impact other dMRI modelling techniques which model the signal as a combination of a diffusion response with an orientation distribution. For instance, NODDI models the intracellular signal with a Watson distribution as the orientation component and diffusion in sticks as the MR response, assuming that all the fibres can be treated as sticks. As shown here, microscopic variations in fibre morphology mean that the signals from each fibre are not identical and this could affect results of NODDI and other similar models (Alexander et al., 2017). Similarly, previous studies have looked into the effect of morphological features such as undulation on axon diameter estimation (Brabec et al., 2019; H.-H. Lee et al., 2020; Nilsson et al., 2012). Future work will aim to shed more light on these effects and investigate whether it is possible for it to be accounted for in our models.

# 5 Conclusion

The complex axonal morphology introduced by axons packing together in complex arrangements leads to differences in the dMRI response across different fibres. These variations in per-fibre response functions can lead to differences in the fODF estimated using CSD when using a single FRF across all fibres and all voxels. Indeed, when the FRF represents the voxel well, CSD can recover the fODF well, however when the FRF is slightly off (as can be the case for assuming a single FRF across all voxels), the fODF can be misestimated.

All of this means that the interpretation of the FRF and fODF in SD needs to be carefully considered and future models may seek to disentangle some of these effects for more accurate FRF and fODF estimation.

## Acknowledgements

This work is supported by the EPSRC-funded UCL Centre for Doctoral Training in Medical Imaging (EP/L016478/1) and the Department of Health's NIHR-funded Biomedical Research Centre at University College London Hospitals. This work was supported by EPSRC grants EP/M020533/1 and EP/N018702/1. MP is supported by UKRI Future Leaders Fellowship MR/T020296/2.

## Author Credit

**Ross Callaghan:** Conceptualisation, Methodology, Software, Investigation, Visualisation, Writing – Original Draft. **Daniel C. Alexander:** Conceptualisation, Writing – Review & Editing, Supervision, Resources, Funding acquisition. **Marco Palombo:** Conceptualisation, Methodology, Writing – Review & Editing, Supervision. **Hui Zhang:** Conceptualisation, Methodology, Writing – Review & Editing, Supervision, Resources, Funding acquisition.

## Data and Code Availability

ConFiG code and simulated data will be made available at https://rcallagh.github.io. EM data is available at https://www.cai2r.net/resources/software/intra-axonal-space-segmented-3d-scanning-electron-microscopy-mouse-brain-genu.

## Conflicts of Interest

The authors confirm that there are no conflicts commercial or financial conflicts of interest affecting this work.

## References

Abdollahzadeh, A., Ilya, B., Jokitalo, E., Tohka, J., Sierra, A., 2019. Automated 3D Axonal Morphometry of White Matter. Sci. Rep. 9, 1–16. https://doi.org/10.1038/s41598-019-42648-2

Alexander, D.C., Barker, G.J., Arridge, S.R., 2002. Detection and modeling of non-Gaussian apparent diffusion coefficient profiles in human brain data. Magn. Reson. Med. 48, 331–340. https://doi.org/10.1002/mrm.10209

Alexander, D.C., Dyrby, T.B., Nilsson, M., Zhang, H., 2017. Imaging brain microstructure with diffusion MRI: Practicality and applications. NMR Biomed. 1–26. https://doi.org/10.1002/nbm.3841

Andersson, M., Kjer, H.M., Rafael-Patino, J., Pacureanu, A., Pakkenberg, B., Thiran, J.-P., Ptito, M., Bech, M., Bjorholm Dahl, A., Andersen Dahl, V., Dyrby, T.B., 2020. Axon morphology is modulated by the local environment and impacts the noninvasive investigation of its structure–function relationship. Proc. Natl. Acad. Sci. 117, 33649–33659. https://doi.org/10.1073/pnas.2012533117

Brabec, J., Lasič, S., Nilsson, M., 2019. Time-dependent diffusion in undulating thin fibers: Impact on axon diameter estimation. NMR Biomed. 1–19. https://doi.org/10.1002/nbm.4187

Brechbühler, Ch., Gerig, G., Kübler, O., 1995. Parametrization of Closed Surfaces for 3-D Shape Description. Comput. Vis. Image Underst. 61, 154–170. https://doi.org/10.1006/cviu.1995.1013

Callaghan, R., Alexander, D.C., Palombo, M., Zhang, H., 2020. ConFiG: Contextual Fibre Growth to generate realistic axonal packing for diffusion MRI simulation. NeuroImage 220, 117107. https://doi.org/10.1016/j.neuroimage.2020.117107

Catani, M., Thiebaut de Schotten, M., 2013. Atlas of Human Brain Connections, Atlas of Human Brain Connections. https://doi.org/10.1093/med/9780199541164.001.0001

Christiaens, D., Sunaert, S., Suetens, P., Maes, F., 2017. Convexity-constrained and nonnegativity-constrained spherical factorization in diffusion-weighted imaging. NeuroImage 146, 507–517. https://doi.org/10.1016/j.neuroimage.2016.10.040

Christiaens, D., Veraart, J., Cordero-Grande, L., Price, A.N., Hutter, J., Hajnal, J.V., Tournier, J.D., 2020. On the need for bundle-specific microstructure kernels in diffusion MRI. NeuroImage 208, 116460–116460. https://doi.org/10.1016/j.neuroimage.2019.116460

Cook, P. a, Bai, Y., Seunarine, K.K., Hall, M.G., Parker, G.J., Alexander, D.C., 2006. Camino: Open-Source Diffusion-MRI Reconstruction and Processing. 14th Sci. Meet. Int. Soc. Magn. Reson. Med. 14, 2759–2759.

Cottaar, M., Bastiani, M., Boddu, N., Glasser, M.F., Haber, S., van Essen, D.C., Sotiropoulos, S.N., Jbabdi, S., 2021. Modelling white matter in gyral blades as a continuous vector field. NeuroImage 227, 117693. https://doi.org/10.1016/j.neuroimage.2020.117693

Dell'Acqua, F., Tournier, J.D., 2019. Modelling white matter with spherical deconvolution: How and why? NMR Biomed. 32, 1–18. https://doi.org/10.1002/nbm.3945

Dhital, B., Reisert, M., Kellner, E., Kiselev, V.G., 2019. Intra-axonal diffusivity in brain white matter. NeuroImage 189, 543–550. https://doi.org/10.1016/j.neuroimage.2019.01.015
Hall, M.G., Alexander, D.C., 2009. Convergence and Parameter Choice for Monte-Carlo Simulations of Diffusion MRI. IEEE Trans. Med. Imaging 28, 1354–1364. https://doi.org/10.1109/TMI.2009.2015756
Jbabdi, S., Johansen-Berg, H., 2011. Tractography: Where Do We Go from Here? Brain Connect. 1, 169–183. https://doi.org/10.1089/brain.2011.0033
Johansen-Berg, H., Behrens, T.E.J., 2006. Just pretty pictures? What diffusion tractography can add in clinical neuroscience. Curr. Opin. Neurol. https://doi.org/10.1097/01.wco.0000236618.82086.01
Kaden, E., Kelm, N.D., Carson, R.P., Does, M.D., Alexander, D.C., 2016a. Multi-compartment microscopic diffusion imaging. NeuroImage 139, 346–359. https://doi.org/10.1016/j.neuroimage.2016.06.002
Kaden, E., Kruggel, F., Alexander, D.C., 2016b. Quantitative mapping of the per-axon diffusion coefficients in brain white matter: Quantitative Mapping of the Per-Axon Diffusion Coefficients. Magn. Reson. Med. 75, 1752–1763. https://doi.org/10.1002/mrm.25734
Lee, H., Papaioannou, A., Kim, S.-L., Novikov, D.S., Fieremans, E., 2020. A time-dependent diffusion MRI signature of axon caliber variations and beading. Commun. Biol. https://doi.org/10.1038/s42003-020-1050-x
Lee, H.-H., Jespersen, S.N., Fieremans, E., Novikov, D.S., 2020. The impact of realistic axonal shape on axon diameter estimation using diffusion MRI. NeuroImage 223, 117228. https://doi.org/10.1016/j.neuroimage.2020.117228
Lee, H.H., Yaros, K., Veraart, J., Pathan, J.L., Xia, F., Sungheon, L., Novikov, D.S., Fieremans, E., 2019. Along-axon diameter variation and axonal orientation dispersion revealed with 3D electron microscopy : implications for quantifying brain white matter microstructure with histology and diffusion MRI. Brain Struct. Funct. 224, 1469–1488. https://doi.org/10.1007/s00429-019-01844-6
Maier-Hein, K.H., Neher, P.F., Houde, J.C., Côté, M.A., Garyfallidis, E., Zhong, J., Chamberland, M., Yeh, F.C., Lin, Y.C., Ji, Q., Reddick, W.E., Glass, J.O., Chen, D.Q., Feng, Y., Gao, C., Wu, Y., Ma, J., Renjie, H., Li, Q., Westin, C.F., Deslauriers-Gauthier, S., González, J.O.O., Paquette, M., St-Jean, S., Girard, G., Rheault, F., Sidhu, J., Tax, C.M.W., Guo, F., Mesri, H.Y., Dávid, S., Froeling, M., Heemskerk, A.M., Leemans, A., Boré, A., Pinsard, B., Bedetti, C., Desrosiers, M., Brambati, S., Doyon, J., Sarica, A., Vasta, R., Cerasa, A., Quattrone, A., Yeatman, J., Khan, A.R., Hodges, W., Alexander, S., Romascano, D., Barakovic, M., Auría, A., Esteban, O., Lemkaddem, A., Thiran, J.P., Cetingul, H.E., Odry, B.L., Mailhe, B., Nadar, M.S., Pizzagalli, F., Prasad, G., Villalon-Reina, J.E., Galvis, J., Thompson, P.M., Requejo, F.D.S., Laguna, P.L., Lacerda, L.M., Barrett, R., Dell'Acqua, F., Catani, M., Petit, L., Caruyer, E., Daducci, A., Dyrby, T.B., Holland-Letz, T., Hilgetag, C.C., Stieltjes, B., Descoteaux, M., 2017. The challenge of mapping the human connectome based on diffusion tractography. Nat. Commun. 8. https://doi.org/10.1038/s41467-017-01285-x
Mardia, K.V., Jupp, P.E., 2008. Directional Statistics, Directional Statistics. John Wiley & Sons, Ltd. https://doi.org/10.1002/9780470316979

Nilsson, M., Lätt, J., Ståhlberg, F., van Westen, D., Hagslätt, H., 2012. The importance of axonal undulation in diffusion MR measurements: A Monte Carlo simulation study. NMR Biomed. 25, 795–805. https://doi.org/10.1002/nbm.1795

Palombo, M., Ianus, A., Guerreri, M., Nunes, D., Alexander, D.C., Shemesh, N., Zhang, H., 2020. SANDI: A compartment-based model for non-invasive apparent soma and neurite imaging by diffusion MRI. NeuroImage 215, 116835–116835. https://doi.org/10.1016/j.neuroimage.2020.116835

Panagiotaki, E., Hall, M.G., Zhang, H., Siow, B., Lythgoe, M.F., Alexander, D.C., 2010. High-Fidelity Meshes from Tissue Samples for Diffusion MRI Simulations, in: MICCAI. pp. 404–411.

Raffelt, D., Tournier, J.D., Rose, S., Ridgway, G.R., Henderson, R., Crozier, S., Salvado, O., Connelly, A., 2012. Apparent Fibre Density: A novel measure for the analysis of diffusion-weighted magnetic resonance images. NeuroImage 59, 3976–3994. https://doi.org/10.1016/j.neuroimage.2011.10.045

Raffelt, D.A., Tournier, J.D., Smith, R.E., Vaughan, D.N., Jackson, G., Ridgway, G.R., Connelly, A., 2017. Investigating white matter fibre density and morphology using fixel-based analysis. NeuroImage 144, 58–73. https://doi.org/10.1016/j.neuroimage.2016.09.029

Saff, E.B., Kuijlaars, A.B.J., 1997. Distributing many points on a sphere. Math. Intell. 19, 5–11. https://doi.org/10.1007/BF03024331

Schilling, K.G., Daducci, A., Maier-Hein, K., Poupon, C., Houde, J.-C., Nath, V., Anderson, A.W., Landman, B.A., Descoteaux, M., 2019a. Challenges in diffusion MRI tractography – Lessons learned from international benchmark competitions. Magn. Reson. Imaging 57, 194–209. https://doi.org/10.1016/j.mri.2018.11.014

Schilling, K.G., Gao, Y., Stepniewska, I., Janve, V., Landman, B.A., Anderson, A.W., 2019b. Histologically derived fiber response functions for diffusion MRI vary across white matter fibers—An ex vivo validation study in the squirrel monkey brain. NMR Biomed. 32, 1–17. https://doi.org/10.1002/nbm.4090

Sotiropoulos, Stamatios N., Jbabdi, S., Xu, J., Andersson, J.L., Moeller, S., Auerbach, E.J., Glasser, M.F., Hernandez, M., Sapiro, G., Jenkinson, M., Feinberg, D.A., Yacoub, E., Lenglet, C., Van Essen, D.C., Ugurbil, K., Behrens, T.E.J., 2013. Advances in diffusion MRI acquisition and processing in the Human Connectome Project. NeuroImage 80, 125–143. https://doi.org/10.1016/j.neuroimage.2013.05.057

Sotiropoulos, S. N., Moeller, S., Jbabdi, S., Xu, J., Andersson, J.L., Auerbach, E.J., Yacoub, E., Feinberg, D., Setsompop, K., Wald, L.L., Behrens, T.E.J., Ugurbil, K., Lenglet, C., 2013. Effects of image reconstruction on fiber orientation mapping from multichannel diffusion MRI: Reducing the noise floor using SENSE. Magn. Reson. Med. 70, 1682–1689. https://doi.org/10.1002/mrm.24623

Thomas, C., Ye, F.Q., Irfanoglu, M.O., Modi, P., Saleem, K.S., Leopold, D.A., Pierpaoli, C., 2014. Anatomical accuracy of brain connections derived from diffusion MRI tractography is inherently limited. Proc. Natl. Acad. Sci. 111, 16574–16579. https://doi.org/10.1073/pnas.1405672111

Tournier, J.D., Calamante, F., Connelly, A., 2007. Robust determination of the fibre orientation distribution in diffusion MRI: Non-negativity constrained super-resolved spherical deconvolution. NeuroImage 35, 1459–1472. https://doi.org/10.1016/j.neuroimage.2007.02.016

Tournier, J.D., Calamante, F., Gadian, D.G., Connelly, A., 2004. Direct estimation of the fiber orientation density function from diffusion-weighted MRI data using spherical deconvolution. NeuroImage 23, 1176–1185. https://doi.org/10.1016/j.neuroimage.2004.07.037

Tournier, J.D., Smith, R., Raffelt, D., Tabbara, R., Dhollander, T., Pietsch, M., Christiaens, D., Jeurissen, B., Yeh, C.H., Connelly, A., 2019. MRtrix3: A fast, flexible and open software framework for medical image processing and visualisation. NeuroImage 202, 116137–116137. https://doi.org/10.1016/j.neuroimage.2019.116137

Van Essen, D.C., Jbabdi, S., Sotiropoulos, S.N., Chen, C., Dikranian, K., Coalson, T., Harwell, J., Behrens, T.E.J., Glasser, M.F., 2014. Mapping Connections in Humans and Non-Human Primates, in: Diffusion MRI. Elsevier, pp. 337–358. https://doi.org/10.1016/B978-0-12-396460-1.00016-0

Van Essen, D.C., Ugurbil, K., Auerbach, E., Barch, D., Behrens, T.E.J., Bucholz, R., Chang, A., Chen, L., Corbetta, M., Curtiss, S.W., Della Penna, S., Feinberg, D., Glasser, M.F., Harel, N., Heath, A.C., Larson-Prior, L., Marcus, D., Michalareas, G., Moeller, S., Oostenveld, R., Petersen, S.E., Prior, F., Schlaggar, B.L., Smith, S.M., Snyder, A.Z., Xu, J., Yacoub, E., 2012. The Human Connectome Project: A data acquisition perspective. NeuroImage 62, 2222–2231. https://doi.org/10.1016/j.neuroimage.2012.02.018

Yeh, C.H., Tournier, J.D., Cho, K.H., Lin, C.P., Calamante, F., Connelly, A., 2010. The effect of finite diffusion gradient pulse duration on fibre orientation estimation in diffusion MRI. NeuroImage 51, 743–751. https://doi.org/10.1016/j.neuroimage.2010.02.041

Zhang, H., Schneider, T., Wheeler-Kingshott, C.A., Alexander, D.C., 2012. NODDI: Practical in vivo neurite orientation dispersion and density imaging of the human brain. NeuroImage 61, 1000–1016. https://doi.org/10.1016/j.neuroimage.2012.03.072